\documentclass[nofootinbib,a4paper,aps,prd,10pt,superscriptaddress,showkeys, twocolumn]{revtex4}

\usepackage[utf8]{inputenc}
\usepackage{graphicx}
\usepackage{graphics}
\usepackage{amsfonts}
\usepackage{amssymb} 
\usepackage{amsmath}
\usepackage{hyperref}
\usepackage{natbib}
\usepackage{float}
\usepackage{dcolumn}% Align table columns on decimal point
\usepackage{bm}% bold math
\usepackage{latexsym,color}

\usepackage{makecell}

\def\nat{Nature}
\def\prl{Phys. Rev. Lett.}

\def\prd{Phys. Rev. D}

\def\mnras{Mon. Not. Roy. Astr. Soc.}
\def\apj{Astrophys. J.}

\def\aap{Astron. Astrophys.}
\def\actaa{Acta Astronomica}

\def\nar{New Astronomy Reviews}
\def\solphys{Solar Physics}

\begin{document}

\title{Adiabatic theory in  Kerr spacetimes}

\author{Kuantay~\surname{Boshkayev}}
\email[]{kuantay@mail.ru}
\affiliation{Al-Farabi Kazakh National University, Al-Farabi av. 71, 050040 Almaty, Kazakhstan.}
\affiliation{National Nanotechnology Laboratory of Open Type,  Almaty 050040, Kazakhstan.}

\author{Gulmira~\surname{Nurbakyt}}
\email[]{gumi-nur@mail.ru}
\affiliation{Al-Farabi Kazakh National University, Al-Farabi av. 71, 050040 Almaty, Kazakhstan.}
\affiliation{International engineering technological university, Al-Farabi av. 93g/5, 050060 Almaty, Kazakhstan.}

\author{Hernando~\surname{Quevedo}}
\email[]{quevedo@nucleares.unam.mx}
\affiliation{Al-Farabi Kazakh National University, Al-Farabi av. 71, 050040 Almaty, Kazakhstan.}
\affiliation{Instituto de Ciencias Nucleares, Universidad Nacional Aut\'onoma de M\'exico, Mexico.}
\affiliation{Dipartimento di Fisica and ICRA, Universit\`a di Roma “La Sapienza”, Roma, Italy.}

\author{Gulnara~\surname{Suliyeva}}
\email[]{sulieva.gulnara0899@gmail.com}
\affiliation{Al-Farabi Kazakh National University, Al-Farabi av. 71, 050040 Almaty, Kazakhstan.}
\affiliation{Fesenkov Astrophysical Institute, Observatory 23, 050020 Almaty, Kazakhstan.}

\author{Abylaikhan~\surname{Tlemissov}}
\email[]{tlemissov-ozzy@mail.ru}
\affiliation{Institute of Physics, Silesian University in Opava, Bezrucovo nam. 13, CZ-74601 Opava, Czech Republic.}

\author{Zhanerke~\surname{Tlemissova}}
\email[]{kalymova.erke@mail.ru}
\affiliation{Institute of Physics, Silesian University in Opava, Bezrucovo nam. 13, CZ-74601 Opava, Czech Republic.}

\author{Anar~\surname{Dalelkhankyzy}}
\email[]{dalelkhankyzy.@gmail.com}
\affiliation{Kazakh National Women's Teacher Training University, Ayteke Bi, 99, 050000 Almaty, Kazakhstan.}
\affiliation{Al-Farabi Kazakh National University, Al-Farabi av. 71, 050040 Almaty, Kazakhstan.}

\author{Aliya~\surname{Taukenova}}
\email[]{aliya\_tauken@mail.ru}
\affiliation{Al-Farabi Kazakh National University, Al-Farabi av. 71, 050040 Almaty, Kazakhstan.}
\affiliation{National Nanotechnology Open Laboratory,  Almaty 050040, Kazakhstan.}

\author{Ainur~\surname{Urazalina}}
\email[]{y.a.a.707@mail.ru}
\affiliation{Al-Farabi Kazakh National University, Al-Farabi av. 71, 050040 Almaty, Kazakhstan.}
\affiliation{National Nanotechnology Open Laboratory,  Almaty 050040, Kazakhstan.}

\author{Zdeněk~\surname{Stuchlík}}
\email[]{zdenek.stuchlik@physics.slu.cz}
\affiliation{Institute of Physics, Silesian University in Opava, Bezrucovo nam. 13, CZ-74601 Opava, Czech Republic.}

\date{\today}
\begin{abstract}
We present the main aspects of the adiabatic theory and show that it can be used to study the motion of test particles in general relativity. The theory is based upon the use of vector elements of the orbits and adiabatic invariants. To prove the applicability of the adiabatic theory in Einstein's gravity, we derive a particular representation of the Kerr metric in harmonic coordinates, which allows us to obtain a general formula for  the perihelion shift of test particles orbiting on the non-equatorial plane of a rotating central object. We show that the principle of superposition is fulfilled for the individual effects of the gravitational source mass and angular momentum up to the second order. We demonstrate that the adiabatic theory, along with its simplicity, leads to correct results, which in the limiting cases correspond to the ones reported in the literature.
\end{abstract}

	
	\maketitle
	
\section{Introduction}
The adiabatic theory introduced in the context of quantum mechanics and physics of nuclear particles is based upon the idea of perturbing the Lagrangian or the Hamiltonian of a system and letting some additional parameters to change slowly. Different aspects of the adiabatic theory have been applied in many branches of physics such as thermodynamics, chemistry, and classical and quantum mechanics \cite{henrard1993adiabatic}.

In the context of classical general relativity, an interesting approach to study the motion of test particles  was proposed by M. Abdildin \cite{1988mtge.book.....A} by using a particular static spacetime metric and the  conceptual framework developed by Fock  \cite{1964tstg.book.....F}. 
In Ref.~\cite{1988mtge.book.....A}, the Fock metric was generalized to include the rotation of the source up to the second order in the angular momentum and its internal structure in the post-Newtonian ($\sim 1/c^2$) approximation, where $c$ is the speed of light in vacuum. This extended Fock metric was originally presented in harmonic coordinates \cite{1964tstg.book.....F,2012PhRvD..86f4043B}, which facilitate the study of the motion of test particles by using vectors associated with the trajectories of the test particles. One of the most important consequences of Abdildin's works was the implementation of the adiabatic theory to study the motion of bodies in general relativity, which drastically simplifies the form of the equations of motion derived previously \cite{1988mtge.book.....A,2006mtge.book.....A}. In this work, we will present the main ideas and physical aspects of the adiabatic theory, emphasizing its applicability and simplicity in quite general problems. To show these advantages explicitly, we will investigate the motion of test particles in the gravitational field of a realistic rotating object.

According to observations, all astrophysical massive and compact objects rotate around some axis. The investigation of the motion of test particles in the gravitational field of such objects allows us to understand their basic properties, to check relativistic effects in a strong field regime, and to study the geometric structure of spacetime around rotating objects.

The solution to the Einstein field equations for a static, spherically symmetric object in vacuum is well-known in the literature as the Schwarzschild metric \cite{1916AbhKP1916..189S}. This solution describes new effects that could not be explained within the classical Newtonian theory of gravity \cite{1973grav.book.....M}.

In 1918, Lense and Thirring derived an approximate external solution that takes into account the rotation of the source up to the first order in the angular momentum \cite{1918PhyZ...19..156L,1975ctf..book.....L}. Rotation defines the difference between static and stationary spacetimes and leads to specific gravitational effects. The most noticeable one is the frame dragging effect near the rotating body which manifests itself as the precession of satellites' orbits and gyroscopes on the non-equatorial plane around the axis of rotation of the central object \cite{2013grsp.book.....O}.

The first exact vacuum solution for a stationary, axisymmetric and asymptotically flat gravitational field was derived by Kerr in 1963 \cite{1963PhRvL..11..237K}. The solution was successfully applied to describe the gravitational field of different mass distributions, starting from stellar mass black holes to supermassive black holes at active galactic nuclei \cite{1983bhwd.book.....S}.

Later, in 1968, Ernst developed a procedure to derive new stationary, axisymmetric and asymptotically flat solutions, based on the Papapetrou line element, through the introduction of a complex potential without directly solving Einstein field equations \cite{1968PhRv..167.1175E}. This method led to novel exact solutions, among which the Kerr solution was the simplest case \cite{2009esef.book.....S}. Despite the fact that the Kerr metric does not have a physically realistic/reasonable interior solution, it is widely used in astronomy and astrophysics to study the physics of black holes and the processes taking place in their vicinity \cite{2007A&A...470..401S,2010ApJ...714..748T,2015ApJ...809..127Z,2020MNRAS.496.1115B, 2021EPJC...81..205B, 2021PhRvD.104h4009B, 2021EPJC...81.1067S,2022ApJ...929...28T}. In the present work, we will focus on the approximate Kerr solution and study the motion of test particles in the framework of the adiabatic theory.

The article is organized as follows. 
In Section  \ref{sec:adiabtheory}, we introduce the basic concepts of the adiabatic theory, emphasizing its applicability to study the motion of test bodies.
In Section \ref{sec:kerrmetric}, we present the Kerr metric in harmonic coordinates, which are convenient for the application of the adiabatic theory.  In Section \ref{sec:method},  we obtain an expression for the perihelion shift within the framework of the  adiabatic theory. Finally, Section \ref{sec:concl} contains the concluding remarks. 
%%%%%%%%%%%%%%%%%%%%%%%%%%%%%%%%%%%%%%%%%%%%%%%%%%%%%%%%%%%%%%%%%%%%%%%%%%%%%%%%%%%%%%%%%%%%%%%%%%%%%%%%%%%%%%%%%%%%%%%%%%%%%%%%%%%%%%%%%%%%%%%%%%%%%%%%%%%%%%%%%%%%%%%%%%%%%%%%%%%%%%%%%%%%%%%%%%%%%%%%%%%%%%%%%%%%%%%%%%%%%%%%%%%%%%%%%%%%%%%%%%%%

\section{Adiabatic theory}\label{sec:adiabtheory}

The adiabatic theory is based on the use of the vector elements of orbits,  asymptotic methods of the theory of nonlinear oscillations, and adiabatic invariants. It represents an alternative method to analyze physical phenomena, without explicitly solving field or motion equations.  

According to the adiabatic theory, the motion of test particles/bodies can be described by a Lagrangian which is essentially the perturbation of a known Lagrangian. Consider, for instance, the Kepler problem for a relativistic particle moving in a central field. Then, the corresponding perturbed Lagrangian function can be expressed as 
\begin{equation}
L=-mc^2+\frac{mv^2}{2}+\frac{Gmm_0}{r}+\frac{1}{c^2}F(\vec{r},\vec{v}),
\end{equation}
where $F$ is the perturbation function. Accordingly, the corresponding Hamilton function is written as
\begin{equation}
H=mc^2+\frac{p^2}{2m}-\frac{Gmm_0}{r}-\frac{1}{c^2}F(\vec{r},\vec{p}),
\end{equation}
where $\vec{p}=\partial L/\partial \vec{v}$ is the momentum of a test particle.

The motion of the test particle can be described by the orbital angular momentum vector $\vec{M}$ and the Laplace-Runge-Lenz vector $\vec{A}$, which are integrals of motion defined as
\begin{eqnarray}
\vec{M}&=&\left[\vec{r}, \vec{p} \right],\\
\vec{A}&=&\left[\frac{\vec{p}}{m},\vec{M} \right]-\frac{Gm_0 m}{r}\vec{r} ,
\end{eqnarray}
where $A=Gm_0 me$ is the magnitude (absolute value) of the Laplace-Runge-Lenz vector, $\vec{r}$ is the radius vector of the test particle, $G$ is the gravitational constant, $m_0$ is the mass of a gravitational source (central object), $m$ is the mass of the test particle, and $e$ is the orbit eccentricity. 

The vectors $\vec{M}$ and $\vec{A}$  characterize the shape and position of the orbit in space. Namely, the vector $\vec{M}$ is directed perpendicularly to the orbit plane and the vector $\vec{A}$ is directed towards the perihelion of the orbit. Thus, one can write the equations of motion in a general form as
\begin{eqnarray}
\frac{d\vec{M}}{dt}&=&\frac{dM}{dt}\vec{e}_M+\left[\vec{\Omega}, \vec{M} \right], \label{eq:M}\\
\frac{d\vec{A}}{dt}&=&\frac{dA}{dt}\vec{e}_A+\left[\vec{\Omega}, \vec{A} \right], \label{eq:A}
\end{eqnarray}
where $\vec{e}_M$ and $\vec{e}_A$ are the unit vectors directed along $\vec{M}$ and $\vec{A}$, respectively, and $\vec{\Omega}$ is the angular velocity of rotation of the ellipse ``as a whole'', which is the sought function in this theory.

{}In Ref~\cite{1988mtge.book.....A}, the problem of two rotating bodies was considered based on Fock's  first approximation metric. Its outcomes allow us to derive the basic formulas for the adiabatic theory in the limiting case when one of the  bodies is much more massive than the other one.  The explicit form of the equations of motion ~\eqref{eq:M} and \eqref{eq:A} for a spinning/rotating test body in the field of a rotating central body can be written  as 
\begin{eqnarray}\
	\frac{d\vec{M}}{dt}&=&-\frac{M S_x \omega_y (1-\sqrt{1-e^2})}{2c^2 aPm^2 (1+\sqrt{1-e^2})}\vec{M}+\left[\vec{\Omega}, \vec{M}\right], \label{eq:Mrot}\\
\frac{d\vec{A}}{dt}&=&\frac{MS_y \omega_x}{2c^2 a^2 m^2 (1+\sqrt{1-e^2})^2}\vec{A}+\left[\vec{\Omega}, \vec{A} \right],\label{eq:Arot}
\end{eqnarray}
where $S_x=I \omega_x$ and $S_y=I \omega_y$ are the $x$ and $y$ components of the proper angular momentum of the test body, respectively, $\omega_x$ and $\omega_y$ are the corresponding components of the angular velocity, $I$ is the moment of inertia of the test body, $a$ is the semi-major axis of the orbit, and $P$ is the semilatus rectum. Further details of the derivation of these and corresponding formulas are given in Ref~\cite{1988mtge.book.....A}. Performing  the scalar product of Eq.~\eqref{eq:Mrot} by $\vec{M}$ and of \eqref{eq:Arot} by $\vec{A}$, we obtain
\begin{eqnarray}
	\frac{dM}{dt}&=&-\frac{M^2 S_x \omega_y (1-\sqrt{1-e^2})}{2c^2 aPm^2 (1+\sqrt{1-e^2})}, \label{eq:Mrot2}\\
	\frac{dA}{dt}&=&\frac{MAS_y \omega_x}{2c^2 a^2 m^2 (1+\sqrt{1-e^2})^2}.\label{eq:Arot2}
\end{eqnarray}
Further, from Eqs. \eqref{eq:Mrot2} and \eqref{eq:Arot2}, one obtains the differential equation
\begin{equation}\label{eq:MArot}
	\frac{dM}{dA}=-\frac{e^2 M}{(1-e^2)A}.
\end{equation}
Since 
\begin{equation}
    A=\alpha e, \qquad \alpha=Gm_0m,
\end{equation}
Eq.~\ref{eq:MArot} can be written as follows
\begin{equation}
   \frac{dM}{M}=-\frac{ede}{(1-e^2)}, 
\end{equation}
which leads to the integral of motion:
\begin{equation}\label{eq:Mconserv}
	\frac{M}{\sqrt{1-e^2}}=const.
\end{equation}
This expression represents and adiabatic invariant of the  system, which can be denoted as
\begin{equation}\label{eq:adinvar}
	M_0 =\frac{M}{\sqrt{1-\frac{A^2}{\alpha^2}}},
\end{equation}
In addition, the adiabatic invariant is related to the nonrelativistic energy as
\begin{equation}\label{eq:energy}
	E=-\frac{Gmm_0}{2a}=-\frac{m\alpha^2}{2M_0^2}.
\end{equation}
By plugging the invariant into \eqref{eq:adinvar}, one can rewrite Eqs.~\eqref{eq:M}, \eqref{eq:A} in a more compact form as
\begin{eqnarray}
	\frac{d\vec{M}}{dt}&=&\frac{dM}{dt}\vec{e}_M+\left[\vec{\Omega}, \vec{M} \right],\label{eq:M2}\\
	\frac{d\vec{e}_A}{dt}&=&\left[\vec{\Omega}, \vec{e}_A \right], \label{eq:A2}
\end{eqnarray}
which can be used to determine the angular velocity $\vec{\Omega}$.
On the other hand, the angular velocity of the orbit as a whole  $\vec{\Omega}$ is found by averaging the equations of motion, which is a procedure that  belongs to the asymptotic methods of nonlinear mechanics. In fact, Abdildin proved that the partial derivative of the averaged Hamiltonian of the system with respect to the orbital angular momentum $\vec{M}$ results in the angular velocity, i.e.,

%\color{red}
%A proof of this statement could be required by the referees. Do we have it? {\bf Yes, in Ref.~\cite{1988mtge.book.....A} page 131 Eq. (23.19) and page 134 Eq. (24.15). Ok, let's leave this part as it is, and later we will complement it if the referees ask so. }
%\color{black}

\begin{equation}\label{eq:omega}
\vec{\Omega}=\frac{\partial \overline{H}}{\partial \vec{M}},
\end{equation}
where  $\overline{H}$ is the Hamiltonian averaged over the period of the test particle's Keplerian orbit. The averaged Hamiltonian depends on the orbital angular momentum $\vec{M}$ and the adiabatic invariant $M_0$ of the system.

The explicit form of $\vec{\Omega}$ depends on the physical system under consideration and is determined by the set of Eqs. \eqref{eq:M2}--\eqref{eq:omega}, which are valid even in the limiting case of a non-spinning test body. 

Thereby, in the adiabatic theory, Eqs.~\eqref{eq:M2} and \eqref{eq:A2} and the expression \eqref{eq:omega} are the mathematical basis for the investigation of the motion of bodies. In other words, these equations completely solve the problem of evolution in the quasi-Kepler problem.

\begin{figure}[ht]
\includegraphics[width=\linewidth]{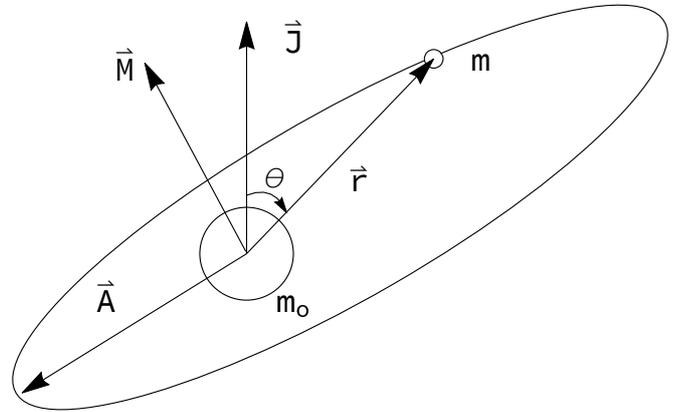} 
\caption{Schematic illustration of a central object and a test particle with its vector elements, where $\theta$ is the polar angle between the proper angular momentum of the source $\vec{J}$ and the radius vector $\vec{r}$}.
\label{fig:pic1}
\end{figure}

In Fig.~\ref{fig:pic1}, we show the position of the vectors elements and the proper angular momentum of the central object $\vec{J}$. 
For further technical purposes, we adopt two reference frames, namely, a fixed frame with coordinates $x_0, y_0, z_0$ and a rotating frame with coordinates $x, y, z$, both having the same origin. In this regard, for generality, the orbital angular momentum of the test particle $\vec{M}$ is directed along the $z$ axis, the Laplace-Runge-Lenz vector is directed along the $x$ axis, the radius vector $\vec{r}$ lies on the $xy$ plane, and the proper angular momentum of the source $\vec{J}$ is directed along the $z_0$ axis. For more details, see Fig.~\ref{fig:pic2}.

%%%%%%%%%%%%%%%%%%%%%%%%%%%%%%%%%%%%%%%%%%%%%%%%%%%%%%%%%%%%%%%%%%%%%%%%%%%%%%%%%%%%%%%%%%%%%%%%%%%%%%%%%%%%%%%%%%%%%%%%%%%%%%%%%%%%%%%%%%%%%%%%%%%%%%%%%%%%%%%%%%%%%%%%%%%%%%%%%%%%%%%%%%%%%%%%%%%%%%%%%%%%%%%%%%%%%%%%%%%%%%%%%%%%%%%%%%%%%%%%%%%%%%%%%%

\section{The Kerr metric in harmonic coordinates}\label{sec:kerrmetric}
	
The form of the Kerr metric in  Boyer-Lindquist coordinates is given by 
\begin{equation}\label{eq:kerr_m}
	\begin{gathered}
		ds^2=\left(1-\frac{2\mu R}{R^2 +a^2 \cos^2\Theta} \right)c^2 dt^2 -\frac{R^2 +a^2 \cos^2\Theta}{R^2-2\mu R +a^2}dR^2\\ -(R^2 +a^2 \cos^2\Theta)d\Theta^2+\frac{4 \mu R a \sin^2\Theta }{R^2 +a^2 \cos^2\Theta}dtd\phi\\-\left(R^2 +a^2 +\frac{2 \mu R a^2 \sin^2\Theta}{R^2 +a^2 \cos^2\Theta}\right)\sin^2\Theta d\phi^2,
	\end{gathered}		
\end{equation}
where $\mu$ and $a$ are related to the mass $m_0$ and angular momentum $J$ of the source as follows
	\begin{equation}
		\begin{gathered}\label{eq:mangmom}
			\mu=\frac{Gm_0}{c^2},\qquad a=\frac{J}{m_0 c},
		\end{gathered}		
	\end{equation}
where $G$ is the gravitational constant and $c$ is the speed of light.

For the purpose of this work, it is convenient to introduce harmonic coordinates $x^\mu$ that satisfy the conditions
\begin{equation}\label{eq:harm_cond}
	\frac{\partial}{\partial x^\nu}\left(\sqrt{-g} g^{\mu \nu} \right)=0,
\end{equation}
where $g^{\mu \nu}$ is the metric tensor in a contra-variant form and $g$ is the determinant of $g_{\mu \nu}$  \cite{1964tstg.book.....F}.
Besides, the harmonic coordinate conditions can be expressed in terms of the Christoffel symbols $\Gamma ^{\mu}_{\ \nu \lambda}$ as
\begin{equation}\label{eq:harm_cond2}
	\Gamma ^{\mu}_{\ \nu \lambda} g^{\nu \lambda}%=\frac{1}{2}g^{\alpha \delta}(g_{\gamma \delta,\beta}+g_{\beta \delta,\gamma}-g_{\beta \gamma,\delta}) g^{\beta \gamma}
	=0,
\end{equation}
which are more convenient from a practical point of view \cite{1993agr..book.....S}.
Harmonic coordinates are important for many problems in general relativity \cite{1964tstg.book.....F}.  For instance, they are related to the conditions under which spacetime can be considered homogeneous and isotropic at large distances from the gravitational field source. In turn, a consequence of the homogeneity and isotropy of the spacetime is the conservation of energy, momentum and angular momentum, which are in fact first integrals of motion.

For the application in adiabatic theory, the Kerr metric \eqref{eq:kerr_m} is expanded in a series of powers of $1/c^2$ as in the post-Newtonian approximation.
Moreover, the following coordinate transformation relates the Boyer-Lindquist coordinates with the harmonic coordinates \cite{bosh2015}
\begin{eqnarray}%\label{eq:transf}
		R&=&r+\frac{1}{c^2}\left(Gm_0-\frac{J^2 \sin^2\theta}{2m_0^2 r}\right), \\ \Theta&=&\theta-\frac{J^2 \sin\theta \cos\theta}{2c^2 m_0^2 r^2}.
\end{eqnarray}
Hence, the approximate Kerr metric in harmonic coordinates reads
\begin{equation}\label{eq:kerr_in_harm}
	\begin{gathered}
		ds^2=\left(1 -\frac{2Gm_0}{c^2 r}+\frac{2G^2 m_0^2}{c^4 r^2}+\frac{2G J^2}{c^4 m_0 r^3} P_2(\cos\theta)\right)c^2 dt^2
	 \\-\left(1+\frac{2Gm_0}{c^2 r}\right)(dr^2+r^2 d\theta^2 +r^2 \sin^2\theta d\phi^2)
	 	\\+\frac{4 G J}{c^2 r}\sin^2\theta dtd\phi\ ,
	\end{gathered}		
\end{equation}
where $P_2(\cos\theta)=(3\cos^2\theta - 1)/2$ is the Legendre polynomial.

The harmonic representation of the Kerr metric \eqref{eq:kerr_in_harm} is essential to explicitly identify relativistic corrections. Thus, in the temporal component of the metric tensor, the first two terms refer to the Newtonian theory and the last two terms are pure relativistic since they are proportional to $1/c^2$.  Moreover, the terms proportional to $1/c^2$ also appear in the spatial and mixed components of the metric tensor.

We have verified that the metric \eqref{eq:kerr_in_harm} fulfills the harmonic coordinate condition in the approximation $\sim 1/c^2$. It also should be mentioned that there are several references in the literature for the Kerr metric in harmonic coordinates, which is derived by employing various mathematical methods \cite{ruiz1986,10.1143/PTP.78.1186,1998ChPhL..15..313L,2001GReGr..33.1809A, 2005CzJPh..55..105B,2014GReGr..46.1671J}. Our representation of the Kerr metric in harmonic coordinates \eqref{eq:kerr_in_harm} is consistent with the one derived in Ref.~\cite{2014GReGr..46.1671J} in the limiting case $\sim 1/c^2$. 

%%%%%%%%%%%%%%%%%%%%%%%%%%%%%%%%%%%%%%%%%%%%%%%%%%%%%%%%%%%%%%%%%%%%%%%%%%%%%%%%%%%%%%%%%%%%%%%%%%%%%%%%%%%%%%%%%%%%%%%%%%%%%%%%%%%%%%%%%%%%%%%%%%%%%%%%%%%%%%%%%%%%%%%%%%%%%%%%%%%%%%%%%%%%%%%%%%%%%%%%

\section{Methodology, results and analyses}\label{sec:method}
As already mentioned, to apply the adiabatic theory in Kerr spacetimes, we need the Kerr metric expanded in powers of $1/c^2$ and written in harmonic coordinates as given in Eq.(\ref{eq:kerr_in_harm}). Then, from the explicit form of the metric one finds the Lagrange function of the test particle
\begin{equation}\label{eq:lagr}
	\begin{gathered}
		L=-mc\frac{ds}{dt}=-mc^2 +\frac{mv^2}{2}+\frac{Gm_0 m}{r}+\frac{mv^4}{8c^2}\\+\frac{3Gm_0 m v^2}{2c^2 r}-\frac{G^2 m_0^2 m}{2c^2 r^2}\\-\frac{GmJ^2}{c^2 m_0 r^3}P_2(\cos\theta)-\frac{4G(\vec{v}\cdot[\vec{r},\vec{J}])}{c^2 r^3}.
	\end{gathered}
\end{equation}
Moreover, the velocity of test particles is defined in the standard form
\begin{equation}
\vec{v}=\frac{d\vec{r}}{dt}, \quad v^2 =\frac{dr^2}{dt^2} +r^2 \left(\frac{d\theta^2}{dt^2} +\sin^2\theta \frac{d\phi^2}{dt^2}\right).
\end{equation}
Notice that only in harmonic and isotropic coordinates, one can use the form of the linear velocity indicated above.

The next step consists in deriving the Hamilton function 
\begin{equation}
H=(\vec{p}\cdot\vec{v})-L,
\end{equation}
where $\vec{p}=\partial L/\partial \vec{v}$ is the generalized momentum  
\begin{equation}\label{eq:genmom}
	\begin{gathered}
\vec{p}=m\vec{v}+\frac{mv^2}{2c^2}\vec{v}+\frac{3Gm_0 m}{c^2 r}\vec{v}-\frac{4G}{c^2 r^3}[\vec{r},\vec{J}],	
    \end{gathered}
\end{equation}
and $\vec{v}$ is the linear velocity
\begin{equation}\label{eq:linvel}
	\begin{gathered}
\vec{v}=\frac{1}{m}\left(\vec{p}-\frac{p^2}{2m^2c^2}\vec{p}-\frac{3Gm_0}{c^2 r}\vec{p}+\frac{4G}{c^2 r^3}[\vec{r},\vec{J}]\right),	
    \end{gathered}
\end{equation}
Taking Eqs.~\eqref{eq:lagr}-\eqref{eq:linvel} into account, the Hamiltonian becomes:
\begin{equation}\label{eq:hamilt}
	\begin{gathered}
		H=mc^2 +\frac{p^2}{2m}-\frac{Gm_0 m}{r} -\frac{p^4}{8c^2 m^3}-\frac{3Gm_0 p^2}{2c^2 m r}\\+\frac{G^2 m_0 ^2 m}{2c^2 r^2}
		+\frac{GmJ^2}{c^2 m_0 r^3}P_2(\cos\theta)+\frac{2G(\vec{p}\cdot[\vec{r}, \vec{J}])}{c^2 r^3}\ .
	\end{gathered}
\end{equation}

Now, according to the adiabatic theory, one should average each term in \eqref{eq:hamilt} over the period $T$ of rotation of the particle. The average of any function $f$ over the period of rotation is defined by means of:
\begin{equation}\label{eq:aver}
\overline{f}=\frac{1}{T}\underset{0}{\overset{T}{\int}} fdt.
\end{equation}
It is convenient to average using the non-relativistic orbital angular momentum $M$ in  polar coordinates \cite{1969mech.book.....L}
\begin{equation}
M=mr^2 \frac{d\phi}{dt},
\end{equation}
which allows us to switch from an integral over $t$ to an integral over $\phi$ for the motion on an elliptical orbit. Here, we use the following solution to the Kepler problem
\begin{equation}
r=\frac{P}{1+e\cos \phi}, \quad 0<\phi\leq2\pi,
\end{equation}
where $e$ is the orbit eccentricity as before, $P$ is the semilatus rectum, and $\phi$ is the polar angle. Therefore, it turns out that
\begin{equation}\label{eq:fphi}
\overline{f}=\frac{1}{T}\underset{0}{\overset{2\pi}{\int }}f(\phi)\frac{dt}{d\phi}d\phi=\frac{m}{TM}\underset{0}{\overset{2\pi}{\int }}f(\phi)r^2 d\phi.%=\frac{1}{2\pi a b}\underset{0}{\overset{2\pi}{\int }}f(\phi)r^2 d\phi.
\end{equation}
In addition, to average the terms in Eq.~\eqref{eq:hamilt} with the radius vector $\vec{r}$ and momentum $\vec{p}=m\vec{v}$, we use the following form of the radius vector on plane and test particle velocity:
\begin{eqnarray} \label{eq:Vvec}
\vec{r}&=&r\left(\vec{i}\cos\phi+\vec{j}\sin\phi\right),\\
\vec{v}&=&\frac{M}{mP}\left(-\vec{i}\sin \phi +\vec{j}(e+\cos \phi)\right).  
\end{eqnarray}
It is also important to mention that one is free to choose the direction of the central body rotation. For simplicity and practical purposes, it is preferred to align it along the $z_0$ axis as $\vec{J}=J\vec{k}_0$. For a test particle moving on the non-equatorial plane, its orbital angular momentum direction does not coincide with the proper angular momentum of the central body (they coincide only on the equatorial plane, where $\theta=\pi/2$).

Then, the averaging is performed in the rotating frame with  coordinates $x, y, z$ over the period of rotation of a particle moving along an elliptical orbit lying on the $xy$ plane.
Applying Eq.~\eqref{eq:aver} or Eq.~\ref{eq:fphi} to each term of the Hamiltonian \eqref{eq:hamilt} and using the formulas for the period, semi-latus rectum and for the eccentricity, expressed in terms of semi-major axis $a$ (not to be confused with the Kerr rotation parameter) \cite{1969mech.book.....L}, respectively,
\begin{equation}\label{eq:TPe}
	\begin{gathered}
 P=\frac{M^2}{m\alpha}=a(1-e^2),\quad e=\sqrt{1-\frac{M^2}{M_0^2}}, \\ 	T=\frac{2\pi M_0^3}{m\alpha^2}=2\pi\sqrt{\frac{ma^3}{\alpha}}, 
	\end{gathered}
\end{equation}

we obtain the averaged Hamilton function:
\begin{equation}
	\begin{gathered}
		\overline{H}=mc^2 -\frac{m \alpha ^2}{2 M_0^2}-\frac{3m\alpha ^4}{c^2 M_0^3 M}+\frac{2m^2 \alpha ^4}{c^2 m_0 M_0^3 M^3}(\vec{J}\cdot \vec{M})\\ +\frac{15 m \alpha ^4}{8c^2 M_0^4}+\frac{m^3 \alpha ^4}{4c^2m_0^2 M_0^3 M^3} \left(J^2-\frac{3(\vec{J}\cdot \vec{M})^2}{M^2}\right)\ .
	\end{gathered}
\end{equation}
As expected, the averaged Hamiltonian depends on the adiabatic invariant $M_0$ and orbital angular momentum $M$.
 
The next step is to find the form of the angular velocity $\vec{\Omega}$. To this end, according to Eq.~\eqref{eq:omega}, we compute the partial derivative of $\overline{H}$ with respect to $\vec{M}$. The result is 
\begin{equation}\label{eq:omegakerr}
	\begin{gathered}
	\vec{\Omega}= \frac{3m\alpha ^4}{c^2 M_0^3 M^2}\vec{e}_M+\frac{2m^2 \alpha ^4 J}{c^2 m_0 M_0^3 M^3}\left(\vec{e}_J-3(\vec{e}_J\cdot \vec{e_M})\vec{e}_M\right)\\- \frac{3m^3 \alpha ^4 J^2}{4 c^2m_0^2 M_0^3 M^4} \left( 2(\vec{e}_J\cdot \vec{e}_M)\vec{e}_J +\left(1-5(\vec{e}_J\cdot \vec{e}_M)^2 \right)\vec{e}_M \right),
	\end{gathered}
\end{equation}
where $\vec{M}=M\vec{e}_M$ and  $\vec{J}=J\vec{e}_J$. The last expression is a generalization of a previous result  obtained in Ref.~\cite{1975ctf..book.....L} for the Lense-Thirring metric, which includes terms $\sim J^2$.

From the general form of the equations of motion \eqref{eq:M2}-\eqref{eq:A2}, it follows that the orbit and the orbital coordinate system associated with it rotates as a rigid body around a fixed point with the angular velocity \eqref{eq:omegakerr}. Therefore, the angular velocity can be associated with the derivatives of the Euler angles, which determine the orientation of the rotating coordinate system relative to the fixed one. 

\begin{figure}[ht]
\includegraphics[width=\linewidth]{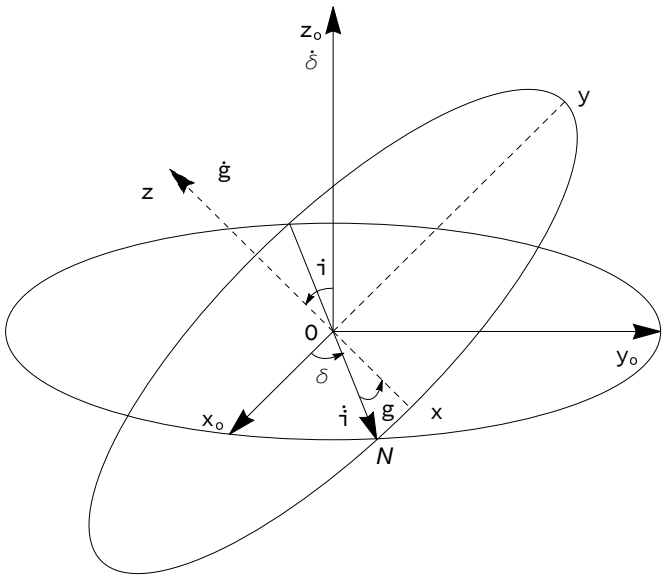} 
\caption{Euler angles and their derivatives.}
\label{fig:pic2}
\end{figure}

Let us denote the fixed coordinate system as $(x_0, y_0, z_0)$ with unit vectors $(\vec{i_0}, \vec{j_0}, \vec{k_0})$ and the rotating one as $(x, y, z)$ with unit vectors $(\vec{i}, \vec{j}, \vec{k})$. 
Moreover, we denote the Euler angles as: $\delta$ is the precession, $g$ is the intrinsic rotation (not to be confused with the determinant of the metric tensor), $i$ is the nutation (inclination angle of the test body orbit).
Then, the angular velocity can be  represented as
\begin{equation}\label{eq:omegaeuler}
\vec{\Omega}=\dot{\delta}\vec{e}_{z_0} +\dot{i}\vec{e}_{\delta} +\dot{g}\vec{e}_z,
\end{equation}
where $\vec{e}_{\delta}$ is the unit vector in the direction of the node line $\vec{N}$, $\vec{e}_{z_0}=\vec{k}_0=\vec{e}_J$ is the unit vector in the $z_0$ direction, $\vec{e}_z=\vec{k}=\vec{e}_M$ is the unit vector in the $z$ direction, and  $\dot{\delta}, \dot{i}, \dot{g}$ are the derivatives of the corresponding Euler angles. It should be mentioned that the vector $\vec{e}_{z}$ is co-linear with $\vec{M}$ and the vector $\vec{e}_{z_0}$ is co-linear with  $\vec{J}$. Then, $i$ represents the angle between $\vec{J}$ and $\vec{M}$ (see Fig.~\ref{fig:pic2}).

Now, considering the above results and comparing the angular velocity \eqref{eq:omegakerr} with the general expression \eqref{eq:omegaeuler}, we obtain
\begin{eqnarray}
\dot{i}&=&0,\label{eq:dot_i}\\
\dot{\delta}&=&\frac{2m^2 \alpha ^4 J}{c^2 m_0 M_0^3 M^3} - \frac{3m^3 \alpha ^4 J^2}{2c^2m_0^2 M_0^3 M^4}(\vec{e}_J\cdot \vec{e}_M),\label{eq:dot_delta}\\
\dot{g}&=&\frac{3m \alpha ^4}{c^2 M_0^3 M^2}-\frac{6 m^2 \alpha ^4  J}{c^2 m_0 M_0^3 M^3}(\vec{e}_J\cdot \vec{e}_M)\nonumber
\\&-&\frac{3m^3 \alpha ^4 J^2}{4c^2m_0^2 M_0^3 M^4}\left(1-5(\vec{e}_J\cdot\vec{e}_M)^2 \right).\label{eq:dot_g}
\end{eqnarray}
From Eq.~\eqref{eq:dot_i}, it follows that the orbit's inclination does not change over time. 

By integrating Eqs.~\eqref{eq:dot_delta}-\eqref{eq:dot_g}, we can find the absolute perihelion shift angle $g_{abs}$, which corresponds to 
\begin{equation}\label{eq:g_abs}
    \Delta g_{abs} = \Delta g + \Delta \delta\ .
\end{equation}
Thereby, we obtain
\begin{equation}\label{eq:g_abs2}
	\begin{gathered}
		\Delta g_{abs}=\frac{6\pi \alpha ^2}{c^2 M^2}+\frac{4 \pi m\alpha ^2 J}{c^2 m_0 M^3}\left(1-3(\vec{e}_J\cdot\vec{e}_M)\right)\\\qquad-\frac{3\pi m^2\alpha ^2 J^2}{2 c^2m_0^2 M^4}\left(1 +2(\vec{e}_J \cdot \vec{e}_M) -5(\vec{e}_J \cdot \vec{e}_M)^2 \right) \ .
    \end{gathered}
\end{equation}	
Thus, to investigate the motion of a test particle on a non-equatorial plane, one only needs to consider an explicit dependence of the perihelion shift on the orbit inclination angle.
%\color{red} 
On the equatorial plane, $\vec{e}_J$ and $ \vec{e}_M$ are parallel so that 
\begin{equation}\label{eq:eqplane}
	\begin{gathered}
		\Delta g_{abs}=\frac{6\pi \alpha ^2}{c^2 M^2}-\frac{8 \pi m\alpha ^2 J}{c^2 m_0 M^3}+\frac{3\pi m^2\alpha ^2 J^2}{ c^2m_0^2 M^4}\ ,
    \end{gathered}
\end{equation}	
or, equivalently,  
\begin{equation}\label{eq:final}
\Delta g_{abs}=\frac{6\pi Gm_0}{c^2 P}-\frac{8 \pi GmJ}{c^2 MP}+\frac{3\pi J^2}{c^2m_0^2 P^2}.
\end{equation}	
%
%\color{black}

From Eq.~\eqref{eq:final}, we can see that for the  problem under consideration the principle of superposition of relativistic effects is valid due to the approximate character of the solution, which is given in terms of the source mass and angular momentum (up to the second order). The first term corresponds to the solution of the Schwarzschild problem (i.e., the term due to the curvature of spacetime caused by the mass of the central body); the second term arises as the frame dragging effect - the Lense-Thirring effect; and the last term is the correction that takes into account the angular momentum up to the second order.

It should be noted that the effect of the perihelion advance in the Schwarzschild problem is associated with the appearance of the orbital momentum $M$ in the Hamiltonian. In classical mechanics, i.e., in the Kepler problem, there is no such dependence and the perihelion remains motionless.

The resulting formula  \eqref{eq:final} reduces in the corresponding limits to the cases already available in the literature (obtained also in harmonic coordinates). So, \eqref{eq:final} in the limit
\begin{itemize}
\item $J=0$ reduces to the Schwarzschild case;
\item $J \neq 0$ (but $J^2 =0$)  reduces to the Lense-Thirring effect;
\item $J \neq 0$ (and $J^2 \neq 0$) reduces to the case of the extended Fock metric.
\end{itemize}
As noted in Ref.~\cite{2012PhRvD..86f4043B}, the angular momentum $J$ specifies higher multipole moments, primarily the quadrupole moment $Q$, which in the case of the extended Fock metric reads $Q=\kappa J^2/(m_0 c^2)$, where different values of $\kappa$ correspond to the following limiting cases (in the $ 1/c^2$ approximation):  
	\begin{itemize}
		\item $\kappa=1$ for the Kerr metric;
		\item $\kappa=4/7$ for the liquid body metric;
		\item $\kappa=15/28$ for the solid body metric.
	\end{itemize}
Besides, it was shown that by means of an  appropriate coordinate transformation, the Kerr solution, expanded to the second order in the angular momentum, transforms into the exterior Hartle-Thorne solution with a particular value of the quadrupole parameter. \footnote{We have to stress that the Kerr approximation can be very useful in many simulations of physical situations connected with rotating objects, but it has a relevant disadvantage as related to the Hartle-Thorne metric. The quadrupole moment of the Kerr metric is not independent from its spin, contrary to the case of the Hartle-Thorne external metric, where these two parameters are independent. For this reason we have to be very careful when treating the predictions of the Kerr approximations, as they could give both insufficient, and false, information in comparison to those related to the Hartle-Thorne external metric, as demonstrated in Ref.~\cite{2021AcA....71..311S}.} Notice that in \cite{1988mtge.book.....A} and  \cite{2006mtge.book.....A} the angular momentum of the central body is denoted by $S_0$ and the motion is limited to the equatorial plane.

%\color{red}
It is interesting to compare the result of this work with the
perihelion/pericenter shift formula 
\begin{equation}
\begin{gathered}
 \Delta\phi=6\pi\frac{M}{r_c}+3\pi \frac{Q}{r_c}\left(\frac{1}{L_z^2}+\frac{2}{r_c^2}\right)-3\pi\frac{J^2}{r_c^2}\left(\frac{4}{L_z^2}+\frac{59}{2r_c^2}\right)\\-8\pi\frac{JM}{L_zr_c}\sqrt{L_z^2+r_c^2}\left(\frac{1}{L_z^2}+\frac{1}{r_c^2}\right)\\
 +24\pi\frac{JQ}{L_zr_c^3}\sqrt{L_z^2+r_c^2}\left(\frac{1}{L_z^2}+\frac{3}{r_c^2}\right)\\+27\pi\frac{M^2}{r_c^2}+3\pi\frac{MQ}{2r_c^2}\left(\frac{30}{L_z^2}+\frac{53}{r_c^2}\right)\\
 +\frac{9}{4}\pi\frac{Q^2}{r_c^2}\left(\frac{3}{L_z^4}+\frac{22}{L_z^2r_c^2}+\frac{63}{r_c^4}\right)\ ,
 \end{gathered}
\end{equation}
which was obtained in 
Ref.~\cite{2019JPhCo...3h5018A}
for the generalized 
 Hartle-Thorne metric and  includes terms 
 proportional to $Q^2$. Thus, 
 to compare with the result obtained in this work for the Kerr metric, we should neglect terms $\sim Q^2$, $\sim M^2$, and  $\sim JQ$. Moreover, the quadrupole moment must be written as $Q=J^2/M$, where $M$ in our notation is the mass of the central body, $M=m_0$, $r_c$ is the semilatus rectum, $r_c=P$, $L_z$ is the orbital angular momentum of a test particle per unit mass and in our notation,  $L_z=M/m$. Then, we obtain 
\begin{equation}
\begin{gathered}
 \Delta\phi
 \approx6\pi\frac{M}{r_c}-8\pi\frac{JM}{L_zr_c}\sqrt{L_z^2+r_c^2}\left(\frac{1}{L_z^2}+\frac{1}{r_c^2}\right)\\+3\pi \frac{J^2}{Mr_c}\left(\frac{1}{L_z^2}+\frac{2}{r_c^2}\right)+3\pi\frac{J^2}{2r_c^2}\left(\frac{11}{L_z^2}-\frac{3}{r_c^2}\right)
 \end{gathered}
\end{equation}
Furthermore, to recover the above formula in  physical units, we introduce the following relationships 
\begin{equation}
    M\rightarrow\frac{Gm_0}{c^2},\quad J\rightarrow\frac{GJ}{c^3}, \quad L_z\rightarrow\frac{M}{c m}, \quad r_c\rightarrow P \ .
\end{equation}
Finally, if we retain only terms $\sim1/c^2$, the final expression becomes
\begin{equation}
\begin{gathered}
 \Delta\phi\approx\frac{6\pi Gm_0}{c^2P}-\frac{8\pi G^2Jm_0m^3}{c^2M^3}+\frac{3\pi G J^2m^2}{c^2m_0 PM^2}
 \end{gathered}
\end{equation}
which is equivalent to Eq.~\eqref{eq:final}, if we use the inverse relationships $M^2=m\alpha P=Gm_0m^2P$. Thus, our result is consistent with the one reported in Ref.~\cite{2019JPhCo...3h5018A}.
%\color{black}

\begin{table*}[ht]
 	\caption{Orbital parameters and perihelion shift  of Mercury \cite{1993tandexpGP..book.....W, 2006LRinRel.100},  Venus, Earth and the S-cluster star S2, S38, S55, and S62.
 	\cite{2017ApJ...837...30G}.
 	All values are calculated for  100 Earth years.
 	}
 	\label{tabular: tab1}
 	\begin{center}
 		\begin{tabular}{cccccccc}
 			\hline
 			\hline
 			\makecell{Objects} &  \makecell{Semi-major axis,\\ $a$ (AU)} &  \makecell{Eccentricity,\\ $e$} & \makecell{Orbit inclination \\ angle, $i~(^{\circ})$} & \makecell{Sidereal period,\\ $T$ (years)} & $\sim m_0$& $\sim J$ & $\sim J^2$ \\
 			\hline
 			 Mercury & 0.3871 & 0.2056 & 3.38  & 0.24 & 43.05'' &  0.0020'' & 7.14  $\cdot 10^{-15}$''\\
 			
 			Venus & 0.7262 & 0.0068 & 3.86  & 0.6151 & 8.61'' & 0.0002''  & 5.13 $\cdot  10^{-17}$''\\
 			 
 			 Earth & 1 & 0.0167 & 7.15  & 1 & 3.83'' &  0.0001'' & 1.31 $\cdot  10^{-17}$''\\
 			
 			  S2 & 970 & 0.8839 & 134.18 & 16.00 & 78.99' & 1.0829' & 7.89 $\cdot  10^{-18}$'\\ 
 			
 			 S38 & 1022 & 0.8201 & 171.1 & 19.20 & 41.73' & 0.6496' & 5.57 $\cdot  10^{-18}$'\\ 
 			
 		 	 S55 & 780 & 0.7209 & 150.1 & 12.80 & 55.92' & 0.7467' & 5.02 $\cdot  10^{-18}$'\\ 
 			
 			S62 & 740 & 0.9760 & 72.76 & 9.90 & 771.81' & 0.9619' & 3.84 $\cdot  10^{-16}$'\\ 
 			\hline
 				\end{tabular}
 			\end{center}
 	\end{table*}

In Table~\ref{tabular: tab1}, we present the orbital parameters of several astrophysical objects and calculate the perihelion shift according to Eq.~\eqref{eq:g_abs2}. All the corrections are evaluated separately to estimate the individual contribution of each parameter. The perihelion shift is calculated for the inner planets of the Solar system and for the S-cluster stars moving in the gravitation field of SgrA* - black hole that is in the center of the Milky Way Galaxy. For the Sun we take the mass $m_0 = M_{\odot}=2 \cdot 10^{30}$ kg and the angular momentum $J = 1.92 \cdot 10^{41}$ kg~m$^2$~s$^{-1}$\cite{2012SoPh..281..815I}. As for the SgrA*, we use $m_0 = 4.2 \cdot 10^6 M_{\odot}$ and $a_{*} = 0.44cJ/Gm_0^2$ \cite{2010MNRAS.403L..74K}, so $J = 6.82 \cdot 10^{54}$ kg~m$^2$~s$^{-1}$. For  the calculations, it is necessary to consider the orbit inclination angle.
{As one can see from the Table, the contribution to the perihelion shift proportional to $J$ is always small with respect to the contribution of $ m_0$. Moreover, the values $\sim J^2$ are always negligible with respect to the values $\sim J$. For the S-cluster stars, which are close to the center of the Milky Way Galaxy, the periastron advance effect is more pronounced in comparison with the Solar system.}
%%%%%%%%%%%%%%%%%%%%%%%%%%%%%%%%%%%%%%%%%%%%%%%%%%%%%%%%%%%%%%%%%%%%%%%%%%%%%%%%%%%%%%%%%%%%%%%%%%%%%%%%%%%%%%%%%%%%%%%%%%%%%%%%%%%%%%%%%%%%%%%%%%%%%%%%%%%%%%%%%%%%%%%%%%%%%%%%%%%%%%%%%%%%%%%%%%%%%%%%%%%%%%%%%5%%%%%
\section{Conclusion}\label{sec:concl}
In this work, we present the fundamentals of the adiabatic theory as developed by Fock and Abdildin. One of the main advantages of this formalism is that it can be applied to classical mechanical and relativistic systems in the same manner. The only input that is needed is the Lagrangian of the corresponding system. Moreover, the investigation of the  motion equations reduces to the analysis of first integrals and adiabatic invariants.
The method is especially adapted to analyze perturbations of a known simple Lagrangian, for which analytical solutions are explicitly known. 

As a particular example of the application of the adiabatic theory in general relativity, we considered the motion of test particles in the gravitational field described by the Kerr metric. To this end, the Kerr metric was expanded in a series in powers of $1/c^2$ and written in harmonic coordinates.
As a result, we derived the perihelion shift expression for test particles moving on the non-equatorial plane of a rotating compact object.  The influence of the central body rotation (up to the second order in the angular momentum) on the test particles trajectory was shown explicitly. The outcomes are obtained by simply considering the dependence of the perihelion shift on the orbit inclination angle. This represents a major advantage in comparison with other procedures known in the literature .

It was also demonstrated that the resulting expression for the perihelion shift satisfies the principle of superposition of relativistic effects due to the approximate character of the solution as given in terms of the source's mass and angular momentum. In the limiting cases, on the equatorial plane, the perihelion shift formula corresponds to the values presented previously in the literature.

It is well known that for the solar system the mass of the Sun is the main parameter. The angular momentum along with the quadrupole moment are negligible in comparison with the effects caused by the solar mass. Nevertheless, the obtained results can be applied to study the motion of stars and planets in the field of a rotating supermassive black holes \cite{2019JPhCo...3h5018A,2019ApJ...886..107W,2021ApJ...909...96W}. In addition, one can also use the results for pulsar planets \cite{1992Natur.355..145W,1994Sci...264..538W,2012NewAR..56....2W}, where the relativistic effects due to the rotation of the central body are more noticeable with respect to our solar system.

%%%%%%%%%%%%%%%%%%%%%%%%%%%%%%%%%%%%%%%%%%%%%%%%%%%%%%%%%%%%%%%%%%%%%%%%%%%%%%%%%%%%%%%%%%%%%%%%%%%%%%%%%%%%%%%%%%%%%%%%%%%%%%%%%%%%%%%%%%%%%%%%%%%%%%%%%%%%%%%%%%%%%%%%%%%%%%%%%%%%%%%%%%%%%%%%%%%%%%%%%%%%%%%%%%%%

\section{Appendix}\label{sec:appendix}

Here we show how the angular velocity of rotation of the ellipse as a whole $\vec{\Omega}$ is determined via the averaged Hamilton function
\begin{equation}\label{eq:harm_cond3}
	\vec{\Omega}=\frac{\partial\overline{H}}{\partial\vec{M}}.
\end{equation}
\subsection{The Lagrange function for two rotating bodies}
Using the Fock approach, the Lagrange function of the rotating N-body problem was derived in Refs.~\cite{1972rcm..book.....B,1988mtge.book.....A}. Both references provide a detailed derivation of the Lagrangian and focus on the problem of two rotating bodies. The main differences between \cite{1972rcm..book.....B} and \cite{1988mtge.book.....A} lies in the assumptions of internal structure of bodies and approximations assumed while computing some integrals. Nevertheless, in the limit of non-rotating bodies, neglecting the internal structure, the Lagrange functions given in Refs.~\cite{1972rcm..book.....B,1988mtge.book.....A} reduce to the ones well-known in the classic literature \cite{1964tstg.book.....F,1975ctf..book.....L}.

Here, our consideration is based on the Lagrange function of a test body with proper rotation in the field of a rotating massive body according to ~\cite{1988mtge.book.....A,2006mtge.book.....A}
  %
% 
%The vector-potential $\vec{U}$ is given as follows 
  %
 % \begin{equation}
%\vec{U}=\frac{G}{2}\left[\vec{\nabla}\frac{1}{r},\vec{J}\right]
 % \end{equation}
 %
\begin{equation}\label{eq:tblag}
L=-m c^2 +m\left(U+\frac{v^2}{2}\right)+T+\delta L
\end{equation}
 where $\delta L$ is the perturbation function
\begin{equation}\label{eq:lagr'}
\begin{gathered}
\delta L=\frac{1}{c^2}\left\{\frac{m v^4}{8}+\left(\frac{\varepsilon}{3}+\frac{3T}{2}\right)v^2-\frac{1}{4}I\left(\vec{\omega}\cdot\vec{v}\right)^2\right.
\\-\frac{m U^2}{2}+\frac{3mv^2 U}{2}+G\left(\left[\vec{S},\vec{\nabla}\right]\cdot\left[\vec{J},\vec{\nabla}\right]\right)\frac{1}{r}
\\+\frac{G}{2}\left(\left[\left(3m_0 \vec{S}+4m\vec{J}\right),\vec{\nabla}\frac{1}{r}\right]\cdot\vec{v}\right)
\\+\left. \frac{2Gm}{7m_0}\left(\left[\vec{J},\vec{\nabla}\right]\cdot\left[\vec{J},\vec{\nabla}\right]\right)\frac{1}{r}\right\},
\end{gathered}
\end{equation}
$\varepsilon$ is the gravitational binding energy taken with opposite sign \cite{1964tstg.book.....F}, $T$ is the rotational kinetic energy, %of the central body, 
$I$ is the moment of inertia relative to the rotation axis of a test body, $U=G m_0/r$ is the Newtonian potential, $\vec{S}$ is the proper angular momentum of the test body, $\vec{J}$ is the proper angular momentum of the central body, and $\vec{\nabla}$ is the nabla operator. It should be noted that this Lagrange function slightly differs from the one presented in Ref.~\cite{1972rcm..book.....B}. Namely, it was assumed that the mass and the size of the test body are much less than the ones of the central body. Then, all terms proportional to the square of the radius of the test body were neglected. However, the term $\frac{2Gm}{7m_0}\left(\left[\vec{J},\vec{\nabla}\right]\cdot\left[\vec{J},\vec{\nabla}\right]\right)\frac{1}{r}$ is the main difference between Ref.~\cite{1988mtge.book.....A} and Ref.~\cite{1972rcm..book.....B}.

\subsection{Equations of motion}

We write the equations of motion in the representation of the vector elements $\vec{M}$ and  $\vec{A}$. This allows one to exploit the asymptotic methods of nonlinear mechanics. In this case, the quantities in the equations of motion can be separated into fast and slow variables. The last circumstance is the distinctive feature of the problems for which the analyses of the asymptotic methods are used. %%[27] 

Thus, the equations of motion are given by
\begin{eqnarray}
\begin{gathered}
\dot{\vec{M}}=\left[\dot{\vec{r}},\vec{p}\right]+\left[\vec{r},\dot{\vec{p}}\right],\\
\dot{\vec{A}}=\left[\frac{\dot{\vec{p}}}{m},\vec{M}\right]+\left[\frac{\vec{p}}{m},\dot{\vec{M}}\right]-m\left[\vec{\nabla} U,\left[\vec{r},\dot{\vec{r}}\right]\right]
\end{gathered}
\end{eqnarray}
The momentum of the test body is written as follows
\begin{equation}
\begin{gathered}
\vec{p}=\frac{\partial L}{\partial\vec{v}}=m\vec{v} +\frac{m\vec{v}}{c^2}\left\{\frac{v^2}{2}+\left(\frac{2\varepsilon}{3m}+\frac{3T}{m}\right)+3U\right\}
\\-\frac{1}{2c^2}\left(\vec{S}\cdot\vec{v}\right)\Vec{\omega}+\frac{G}{2c^2}\left[\left(3m_0 \vec{S}+4m\vec{J}\right),\vec{\nabla}\frac{1}{r}\right]
\end{gathered}
\end{equation}
The corresponding Hamilton function is given by
\begin{equation}
H=mc^2+\frac{p^2}{2m}+\frac{{S}^2}{2I}-m U+\delta H,
\end{equation}
where $\delta H$
\begin{equation}\label{eq:hamil_2}
\begin{gathered}
\delta H=-\frac{1}{c^2}\left\{\frac{p^4}{8m^3}+\left(\frac{\varepsilon}{3}+\frac{3T}{2}\right)\frac{p^2}{m^2}-\frac{I\left(\vec{\omega}\cdot\vec{p}\right)^2}{4m^2}\right.
\\ +\frac{3p^2U}{2m}-\frac{mU^2}{2}+G\left(\left[\vec{S},\vec{\nabla}\right]\cdot\left[\vec{J},\vec{\nabla}\right]\right)\frac{1}{r}
\\+\frac{G}{2m}\left(\left[\left(3m_0 \vec{S}+4m\vec{J}\right),\vec{\nabla}\frac{1}{r}\right]\cdot\vec{p}\right)
\\\left.+\frac{2Gm}{7m_0} \left(\left[\vec{J},\vec{\nabla}\right]\cdot\left[\vec{J},\vec{\nabla}\right]\right)\frac{1}{r}\right\}
\end{gathered}
\end{equation}

As one may notice, the expressions \eqref{eq:lagr'} and \eqref{eq:hamil_2}, within the approximation of $\sim 1/c^2$, are related as
\begin{equation}
\delta H=-\delta L.
\end{equation}
We will come back to this relation later.  

Now, using the canonical equations
\begin{eqnarray}
\dot{\vec{r}}&=&\frac{\partial{H}}{\partial\vec{p}},\\
\dot{\vec{p}}&=&-\frac{\partial{H}}{\partial\vec{r}},  
\end{eqnarray}
one can calculate the derivative of the orbital angular momentum with respect to time,
\begin{equation}\label{eq:vecM}
\begin{gathered}
\dot{\vec{M}}=\frac{1}{2c^2}\left(\Vec{S}\cdot\Vec{v}\right)\left[\vec{\omega},\Vec{v}\right]+\frac{2G}{r^3c^2}\left[\Vec{J},\Vec{M}\right]
\\+\frac{3U}{2mr^2c^2}\left[\Vec{S},\Vec{M}\right]-\frac{3G}{r^5c^2}\left(\Vec{{J}^*}\cdot\Vec{r}\right)\left[\Vec{r},\Vec{J}\right]
\\-\frac{3G}{r^5c^2}\left(\Vec{J}\cdot\Vec{r}\right)\left[\Vec{r},{\Vec{J}}^*\right],
\end{gathered}
\end{equation}
and the derivative of the Laplace-Runge-Lenz vector
\begin{equation}\label{eq:vecA}
\begin{gathered}
\dot{\vec{A}}=\left\{4E+6mU+\left(\frac{2\varepsilon}{3}+3T\right)\right\}\frac{\left[\Vec{\nabla} U,\vec{M}\right]}{mc^2}
\\+\frac{1}{2c^2}\left(\Vec{S}\cdot\Vec{v}\right)\left[\Vec{v},\left[\vec{\omega},\Vec{v}\right]\right]+\frac{1}{2c^2}\left(\Vec{S}\cdot\Vec{v}\right)\left[\Vec{\nabla}U,\left[\vec{\omega},\Vec{r}\right]\right]
\\+\frac{3U}{2m r^2 c^2}\left[\Vec{S},\Vec{A}\right]+\frac{6G}{m r^5c^2}\left(\Vec{J}\cdot\Vec{M}\right)\left[\Vec{r},\Vec{M}\right]
\\+\frac{2G}{r^3 c^2}\left[\Vec{J},\Vec{A}\right]+\frac{9U}{2m^2 r^4 c^2}\left(\Vec{S}\cdot\Vec{M}\right)\left[\Vec{r},\Vec{M}\right]
\\-\frac{3G}{m r^5 c^2}\left\{\left(\Vec{J}^*\cdot\Vec{J}\right)\left[\Vec{r},\Vec{M}\right]+\left({\Vec{J}}^*\cdot\Vec{r}\right)\left[\Vec{J},\Vec{M}\right]\right.
\\\left.-\frac{5}{r^2}\left({\Vec{J}}^* \cdot\Vec{r}\right)\left(\Vec{J}\cdot\Vec{r}\right)\left[\Vec{r},\Vec{M}\right]+\left(\Vec{J}\cdot\Vec{r}\right)\left[{\Vec{J}}^* ,\Vec{M}\right]\right.
\\\left.+\left({\Vec{J}}^*\cdot\Vec{r}\right)\left[\Vec{p},\left[\vec{r},\Vec{J}\right]\right]+\left(\Vec{J}\cdot\Vec{r}\right)\left[\Vec{p},\left[\vec{r},{\Vec{J}}^*\right]\right]\right\}
\end{gathered}
\end{equation}
where, for brevity, the following notation 
\begin{equation}
{\Vec{J}}^*=\Vec{S}+\frac{2}{7}\frac{m}{m_0}\vec{J}
\end{equation}
 was introduced.

\subsection{Averaging the equations of motion}

The exact integration of Eqs.~\ref{eq:vecM}-\ref{eq:vecA} is cumbersome. The fact that all computations are performed within the approximation $\sim 1/c^2$ allows one to employ the well-corroborated methods of finding approximate solutions of differential equations, containing a small parameter. In this case, the methods of non-linear mechanics are particularly effective for this purpose.

Differential equations in the first approximation of the asymptotic method are obtained by averaging the right hand sides of Eqs.~\ref{eq:vecM}-\ref{eq:vecA} for slow variables with respect to fast variables. In addition, when averaging all the expressions the Kepler's (non perturbed) values are used, namely  formulas from \eqref{eq:aver} to \eqref{eq:Vvec} are involved.
Thus, after integration and some algebraic manipulations and simplifications, the averaged equations of motion \eqref{eq:vecM} and \eqref{eq:vecA} become
\begin{eqnarray}
	\overline{\frac{d\vec{M}}{dt}}&=&-\frac{M S_1 \omega_2\left(1-\sqrt{1-e^2}\right)} {2a P m^2 c^2\left(1+\sqrt{1-e^2}\right)}\Vec{M}+\left[\vec{\Omega}_M,\vec{M} \right],\label{eq:M3}\qquad\\
	\overline{\frac{d\vec{A}}{dt}}&=&\frac{M S_2 \omega_1} {2a^2 m^2 c^2\left(1+\sqrt{1-e^2}\right)^2}\Vec{A}+\left[\vec{\Omega}_A,\vec{A} \right], \label{eq:A3}
\end{eqnarray}
where
\begin{equation}\label{eq:omM}
\begin{gathered}
\vec{\Omega}_M=\frac{\left(\Vec{M}\cdot\Vec{S}\right)} {2 a b m^2 c^2\left(1+\sqrt{1-e^2}\right)}\left(\frac{\omega_1}{\sqrt{1-e^2}}\vec{i}+\omega_2 \Vec{j}\right)
\\+\frac{G}{a b P c^2}\left\{2\Vec{J}+\frac{3m_0}{2m}\Vec{S}
-\frac{3}{2M^2}\left({\Vec{J}}^*\cdot\vec{M}\right)\vec{J}\right.
\\\left.-\frac{3}{2M^2}\left(\vec{J}\cdot\vec{M}\right){\Vec{J}}^*\right\}
\end{gathered}
\end{equation}
\begin{equation}\label{eq:omA}
\begin{gathered}
\vec{\Omega}_A=\frac{\left(\omega_2 S_2 -\omega_1 S_1\right)} {4 a b m^2 c^2\left(1+\sqrt{1-e^2}\right)^2}\Vec{M} +\frac{3G m_0 \vec{M}} { a b P m c^2}
\\-\frac{\omega_2\left(\Vec{M}\cdot\Vec{S}\right)\vec{j}} {2 a b m^2 c^2\left(1+\sqrt{1-e^2}\right)}+\frac{G}{a b P c^2}\left\{2\Vec{J}\right.
\\-\frac{3}{2M^2}\left(\left({\vec{J}}^*\cdot\vec{M}\right)\vec{J}+\left(\vec{J}\cdot\vec{M}\right){\vec{J}}^*\right)
\\+\left.\frac{3m_0}{2m}\Vec{S}+\frac{3}{M^4}\left({\vec{J}}^*\cdot\vec{M}\right)\left(\vec{J}\cdot\vec{M}\right)\vec{M}\right\}
\\-\frac{3G \vec{M}} {a b P M^2 c^2}\left\{2\left(\Vec{J}\cdot\Vec{M}\right)\right.+\frac{3m_0}{2m}\left(\Vec{S}\cdot\Vec{M}\right)
\\+\left.\frac{1}{2}\left({\vec{J}}^*\cdot\Vec{J}\right)
-\frac{3}{2 M^2}\left({\vec{J}}^*\cdot \vec{M}\right)\left(\vec{J}\cdot \vec{M}\right)\right\}
\end{gathered}
\end{equation}

The resulting equations of motion can be further transformed. Instead of the two angular velocities, one can introduce the angular velocity common to both equations \eqref{eq:omM} and \eqref{eq:omA}
\begin{equation}
%\begin{gathered}
\vec{\Omega}=\vec{\Omega}_{A}-\frac{\left(\Vec{M}\cdot\Vec{S}\right)\omega_1 \vec{i}}{2 a P m^2 c^2\left(1+\sqrt{1-e^2}\right)}
\end{equation}
Then, the equations of motion \eqref{eq:M3} and \eqref{eq:A3} are written as
\begin{eqnarray}
	\frac{d\vec{M}}{dt}&=&-\frac{ S_1\omega_2 M\left(1-\sqrt{1-e^2}\right)} {2 c^2 a P m^2\left(1+\sqrt{1-e^2}\right)}\Vec{M}+\left[\vec{\Omega},\vec{M} \right],\label{eq:M4}\\
	\frac{d\vec{A}}{dt}&=&\frac{S_2\omega_1 M}{2 c^2 a^2 m^2 \left(1+\sqrt{1-e^2}\right)^2}\Vec{A}+\left[\vec{\Omega},\vec{A} \right]. \label{eq:A4}
\end{eqnarray}

These equations are the ones given in Eqs.~\eqref{eq:Mrot}-\eqref{eq:Arot}. Here, for simplicity, the overline (notation of averaging) is omitted.

\subsection{Transformation of the equations of motion}

According to the invariant Eq.~\eqref{eq:adinvar}, the number of independent variables in the first approximation equations can be reduced by one. We take $\vec{A}$ as such a variable. Therefore, \eqref{eq:M4} and \eqref{eq:A4} will contain only the variables $\vec{M}$ and ${\vec{e}}_A$, as well as the adiabatic invariant $M_0$. Consequently,
\begin{eqnarray}
\frac{d\vec{M}}{dt}&=&-\frac{S_1 \omega_2 \alpha^2\left(M_0 -M\right)} {2c^2{M_0}^2\left(M_0 +M\right)}\Vec{e}_M +\left[\vec{\Omega},\vec{M} \right]\\
\frac{d\vec{A}}{dt}&=&\frac{S_2 \omega_1 \alpha^3 M\left({M_0}^2 -M^2\right)^{\frac{1}{2}}} {2c^2{M_0}^3\left(M_0 +M\right)^2}\Vec{e}_A +\left[\vec{\Omega},\vec{A} \right], \qquad
\end{eqnarray}
where $\vec{\Omega}$ is also expressed in terms of the adiabatic invariant $M_0$ and constant $\alpha$.

One can show that $\vec{\Omega}$ can be also expressed via the operator $\frac{\partial}{\partial\vec{M}}$ in terms of a new function $\Pi$
\begin{equation}
\begin{gathered} \label{eq:OM2}
\vec{\Omega}=\frac{\partial\Pi}{\partial\vec{M}}=\frac{1}{c^2}\frac{\partial}{\partial\vec{M}}\left\{-\frac{3m \alpha^4}{{M_0}^3 M}+\frac{I \alpha^2}{4M_0 \left(M_0 +M\right)}\right.
\\\times\left(\frac{1}{M^2}\left(\Vec{M}\cdot\left[\vec{\omega},\vec{j}\right]\right)^2+\frac{1}{M M_0}\left(\vec{M}\cdot\left[\vec{i},\vec{\omega}\right]\right)^2\right)
\\+\frac{m^2 \alpha^4}{m_0 {M_0}^3 M^3}\left[2\left(\vec{J}\cdot\vec{M}\right)+\frac{3 m_0}{2 m}\left({\Vec{S}}\cdot\vec{M}\right)\right.
\\+\left.\left.\frac{1}{2}\left({\Vec{J}}^*\cdot\Vec{J}\right)-\frac{3}{2M^2}\left({\Vec{J}}^*\cdot\vec{M}\right)\left(\vec{J}\cdot \vec{M}\right)\right]\right\}.
\end{gathered}
\end{equation}
Thus, as a result of our transformations we get the equations in the first approximation as follows
\begin{eqnarray}
	\frac{d\vec{M}}{dt}&=&-\frac{ S_1 \omega_2 \alpha^2\left(M_0 -M\right)} {2c^2{M_0}^2 \left(M_0 +M\right)}\Vec{e}_M +\left[\vec{\Omega},\vec{M} \right]\\
	\frac{d\vec{e}_A}{dt}&=& \left[\vec{\Omega},\vec{e}_A \right]
\end{eqnarray}
It should be noted that $\vec{M}$ and $M_0$ have dimensions of action, and $\vec{e}_A$ is dimensionless.

\subsection{Perturbation function and its average}
Regarding the perturbation functions $\delta H= -\delta L$, one can calculate their average values by the Keplerian ellipse and show that their derivatives with respect to  $\vec{M}$  are equal to 

\begin{eqnarray}
\vec{\Omega}=-\frac{\partial\overline{\delta L}}{\partial\vec{M}}=\frac{\partial\overline{\delta H}}{\partial\vec{M}}=\frac{\partial\Pi}{\partial\vec{M}}
\end{eqnarray}
As a result of such transformations and calculations, we see that the angular velocity $\vec{\Omega}$ can be determined using the averaged Hamiltonian $\overline{H}$.

\section{Acknowledgments}\label{sec:acknowledgments} 
KB, AU and AT acknowledge the Ministry of Education and Science of the Republic of Kazakhstan, Grant: IRN AP19680128.
The work of HQ was partially supported  by UNAM-DGAPA-PAPIIT, Grant No. 114520, and  Conacyt-Mexico, Grant No. A1-S-31269.

\end{document}